\documentclass[preprint,superscriptaddress,aps]{revtex4}%
\usepackage{amsfonts}
\usepackage{amsmath}
\usepackage{amssymb}
\usepackage{graphicx}%
\providecommand{\U}[1]{\protect\rule{.1in}{.1in}}

\begin{document}
\title{Foundations for Cooperating with Control Noise \\in the Manipulation of Quantum Dynamics}
\author{Feng Shuang}
\affiliation{Department of Chemistry, Princeton University, Princeton, New Jersey 08544}
\author{Herschel Rabitz}
\affiliation{Department of Chemistry, Princeton University, Princeton, New Jersey 08544}
\author{Mark Dykman}
\affiliation{Department of Physics and Astronomy, Michigan State University, East Lansing,
Michigan 48824}
\date{\today}

\begin{abstract}
This paper develops the theoretical foundations for the ability of a control
field to cooperate with noise in the manipulation of quantum dynamics. The
noise enters as run-to-run variations in the control amplitudes, phases and
frequencies with the observation being an ensemble average over many runs as
is commonly done in the laboratory. Weak field perturbation theory is
developed to show that noise in the amplitude and frequency components of the
control field can enhance the process of population transfer in a multilevel
ladder system. The analytical results in this paper support the point that
under suitable conditions an optimal field can cooperate with noise to improve
the control outcome.

\end{abstract}
\maketitle

\section{Introduction}

\bigskip Control of quantum processes is actively being pursued theoretically
\cite{Rice00,Rabitz041998} and experimentally\cite{Brixner011,Walmsley0343}.
In practice, control field noise and environmental interactions inevitably are
involved. The present paper considers the influence of field noise upon the
controlled dynamics with the noise described by shot-to-shot pulse variations,
as in typical signal averaged experiments. Recent studies considered several
aspects of the influence of laser noise\cite{Malinovsky911538,
Walser92468,Akramine984892,JM0010841,Rabitz0263405,Ignacio049009,
Shuang049270}, and this work aims to further explore the issue. The
interaction between the noise and field driven dynamics is generally a highly
nonlinear process. The impact of the noise can either be
constructive\cite{Grossmann93229,Blanchard94749,Shao9781,Klappauf981203} or
destructive\cite{Ignacio049009} in the manipulation of quantum dynamics as
well as either improve\cite{Gross924557, Toth943715} or reduce\cite{Rana96198}
the convergence rate of control search efforts. These disparate behaviors make
it difficult to precisely identify the role of noise under various
circumstances, but the many successful experiments at least support the point
that some noise can be
tolerated\cite{Gerber98919,Kunde00924,Zeidler0123420,Brixner0157,Levis01709,Daniel03536}%
. Operating under closed-loop\cite{Judson921500} in the laboratory will
naturally deal with noise as best as possible. A theoretical
analysis\cite{Rabitz0263405} on the impact of field noise upon optimal control
indicated that an inherent degree of robustness can be anticipated by virtue
of the controlled observable expectation values being bilinear in the
evolution operator and its adjoint. Some simulations of closed-loop
experiments also show that robust control is possible and that properly
designed control fields can fight against
noise\cite{JM0010841,Gross924557,Toth943715, Shuang049270}.

Recent numerical simulations\cite{Shuang049270} of closed-loop control in a
model system showed that under suitable circumstance the control field could
cooperate with the presence of noise to more efficiently reach the target
state. The cooperation was remarkable when seeking a modest yield (e.g.,
$\sim10\%$) in the target state. In this case the noise or the deterministic
field acting alone would each produced a small yield, but the two acting
together cooperatively produced a much larger yield. The latter numerical
simulations did not reveal the underlying physical origin of the cooperative
effect, and the present paper will analyze the controlled dynamics of a
multistate system in several limiting cases to explain the prior findings. The
cooperative behavior also has foundation in analogous stochastic resonance
phenomena\cite{Dykman95661, Wiesenfeld98539, Shuang007192} and fluctuation
control\cite{Mark972516, Vugmeister972522}.

Section II presents a general control model of population transfer in
multilevel systems. The noise is modeled by run-to-run variations in the
control amplitudes, phases and frequencies with the observation being an
ensemble average over many runs as is commonly done in the laboratory. The
goal of the control is to maximally populate a highly excited state. In Sec.
III we develop a weak field perturbation theory to provide an analytical
solution to the outcome of the controlled dynamics. We obtain the
noise-average yield from applying the control field in Sec. IV. It is shown
that variation of the field phases from pulse-to-pulse plays no role in the
dynamics but strong cooperation between the deterministic portion of the field
and noise, in both the amplitude and frequency, is possible. Finally, we draw
some conclusions and discuss general multistate systems in Sec. V.

\section{The Model System}

The effects of field noise on controlled quantum dynamics will be explored in
the context of population transfer in multilevel systems characterized by the
Hamiltonian $H$,%
\begin{subequations}
\begin{align}
H  &  =H_{0}-\mu E(t)\text{,}\label{Ht}\\
H_{0}  &  =\sum_{n}\varepsilon_{n}\left\vert n\right\rangle \left\langle
n\right\vert \text{,} \label{H0}%
\end{align}
where $\left\vert n\right\rangle $ is an eigenstate of $H_{0}$ with the
associated energy $\varepsilon_{n}$ in the absence of radiation, and $\mu$ is
the dipole operator, $\mu=\sum_{n,n^{\prime}}\mu_{nn^{\prime}}\left\vert
n\right\rangle \left\langle n^{\prime}\right\vert $. The control field $E(t)$
has the form which may be implemented in the laboratory\cite{Rabitz05013419},%
\end{subequations}
\begin{equation}
E(t)=2s(t)\sum_{l=1}^{M}A_{l}\cos\left(  \omega_{l}t+\theta_{l}\right)
\text{,} \label{E0}%
\end{equation}
where $\left\{  \omega_{l}\right\}  $ are the frequencies of the radiation,
and $s(t)$ is the pulse envelope function. The controls are the amplitudes
$\left\{  A_{l}\right\}  $ and phases $\left\{  \theta_{l}\right\}  $.

\qquad Noise in the laboratory could take on various forms and arise from a
number of sources\cite{Ignacio049009}. In keeping with laboratory practice,
the achieved control will be measured as an ensemble average over the outcome
of many noise contaminated control fields, and the noise is modeled as
shot-to-shot uncertainties in the amplitudes $A_{l}$, phases $\theta_{l}$ and
frequencies $\omega_{l}$ in Eq. (\ref{E0})\cite{JM0010841,Shuang049270}:
\begin{subequations}
\label{APFNoise}%
\begin{align}
A_{l}  &  =A_{l}^{0}+\tilde{x}_{l}\text{,}\label{ANoise}\\
\theta_{l}  &  =\theta_{l}^{0}+\tilde{y}_{l}\text{,}\label{PNoise}\\
\omega_{l}  &  =\omega_{l}^{0}+\tilde{z}_{l}\text{,} \label{FNoise}%
\end{align}
with $\left\langle \tilde{x}_{l}\right\rangle =\left\langle \tilde{y}%
_{l}\right\rangle =\left\langle \tilde{z}_{l}\right\rangle =0$. The
shot-to-shot field noises are captured in terms of zero-mean uncertainties
$\vec{x}=\left\{  \tilde{x}_{l}\right\}  $, $\vec{y}=\left\{  \tilde{y}%
_{l}\right\}  $, and $\vec{z}=\left\{  \tilde{z}_{l}\right\}  $. For
simplicity we assume that the different noise components are independent with
distributions $\rho_{l}^{\left(  A\right)  }\left(  \tilde{x}_{l}\right)  $,
$\rho_{l}^{\left(  \theta\right)  }\left(  \tilde{y}_{l}\right)  $ and
$\rho_{l}^{\left(  \omega\right)  }\left(  \tilde{z}_{l}\right)  $,
respectively. In practice the field noise in $E\left(  t\right)  $ may have
complex structures and origins. Besides the uncertainties in the control field
amplitudes $\left\{  A_{l}\right\}  $ and phases $\left\{  \theta_{l}\right\}
$, a potential source of additional uncertainties is in the frequencies
$\left\{  \omega_{l}\right\}  $ due to laser frequency jitter from variations
of the refractive index\cite{Boller912593}, or other sources\cite{Balle9333}.
The flexible treatment of random variations in $\left\{  A_{l}\right\}  $,
$\left\{  \theta_{l}\right\}  $ and $\left\{  \omega_{l}\right\}  $ is meant
to represent all of these various possibilities. The net outcome of the
control experiments is the average,%
\end{subequations}
\begin{equation}
\bar{O}\left[  E\left(  t\right)  ,\vec{\gamma}\right]  =\left(  \prod_{l}%
\int_{-\gamma_{l}}^{\gamma_{l}}\rho_{l}^{\left(  A\right)  }\left(  \tilde
{x}_{l}\right)  d\tilde{x}_{l}\int_{-\gamma_{l}^{\prime}}^{\gamma_{l}^{\prime
}}\rho_{l}^{\left(  \theta\right)  }\left(  \tilde{y}_{l}\right)  d\tilde
{y}_{l}\int_{-\gamma_{l}^{\prime\prime}}^{\gamma_{l}^{\prime\prime}}\rho
_{l}^{\left(  \omega\right)  }\left(  \tilde{z}_{l}\right)  d\tilde{z}%
_{l}\right)  O\left[  E(t,\vec{x},\vec{y},\vec{z})\right]  \text{,}
\label{ONoise}%
\end{equation}
where $O\left[  E(t,\vec{x},\vec{y},\vec{z})\right]  $ is the control yield
\begin{equation}
O\left[  E\left(  t,\vec{x},\vec{y},\vec{z}\right)  \right]  =\left\vert
\left\langle \Psi_{f}|\psi\left[  E(t,\vec{x},\vec{y},\vec{z}),T\right]
\right\rangle \right\vert ^{2}\text{,}%
\end{equation}
produced by the field $E\left(  t,\vec{x},\vec{y},\vec{z}\right)  $ in Eq.
(\ref{E0}) using the amplitudes, phases and frequencies in Eq. (\ref{APFNoise}%
). The target state is $\left\vert \Psi_{f}\right\rangle $, and $\left\vert
\psi\left[  E(t,\vec{x},\vec{y},\vec{z}),T\right]  \right\rangle $ is the
state of the field-driven system at the final time $T$, which is a functional
over time of $E(t,\vec{x},\vec{y},\vec{z})$, $t\leq T$. In what follows we
assume $E\left(  t\right)  \rightarrow0$ for $t\rightarrow\pm\infty$.

The objective function to be minimized with respect to $\left\{  A_{l}%
^{0}\right\}  $ and $\left\{  \theta_{l}^{0}\right\}  $ in the presence of
noise has the form%
\begin{subequations}
\begin{align}
J  &  =\left\vert \bar{O}\left[  E(t),\vec{\gamma}\right]  -O_{T}\right\vert
^{2}+\alpha F_{0}\text{,}\label{JN}\\
F_{0}  &  =\sum_{l}\left(  A_{l}^{0}\right)  ^{2}\text{,} \label{F0}%
\end{align}
where $O_{T}$ is the target value, and $F_{0}$ is the fluence of the control
field whose contribution is weighted by the constant, $\alpha>0$.

\section{Weak field Perturbation Theory}

To illustrate the principle of how the deterministic portion of the control
field can cooperate with the noise, we consider the excitation along a ladder
(or chain) of nondegenerate transitions and energy levels with each linked
only to its nearest neighbors. One could analogously think of the system as a
nonlinear oscillator\cite{Larsen76254, Wallraff04431,Dykman05140508} or a spin
with $S>1$ and nonequidistant energy levels. The transition elements are taken
to have the form%
\end{subequations}
\begin{equation}
\mu_{nn^{\prime}}=\mu_{n}\delta_{n^{\prime}+1,n}+\mu_{n^{\prime}}%
\delta_{n^{\prime},n+1}\text{.}%
\end{equation}
The $N+1$ level system consists of an initially occupied ground state
$\left\vert 0\right\rangle $ at $t\rightarrow-\infty$, $N-1$ intermediate
states $\left\vert n\right\rangle $, $n=1$, $2$, $\cdots$, $N-1$, and a final
target state $\left\vert N\right\rangle $. The states are coupled with an
external laser pulse having the nominal form of Eq. (\ref{E0}). The wave
function is expanded in the form
\begin{equation}
\psi(t)=\sum_{n=0}^{N}C_{n}\left(  t\right)  |n\rangle e^{-i\varepsilon_{n}%
t}\text{.} \label{psi}%
\end{equation}
The initial condition at $t\rightarrow-\infty$ specifies that $C_{0}=1$ and
$C_{n}=0$ for $0<n\leq N$. The goal is to maximize $\left\vert C_{N}\left(
t\right)  \right\vert ^{2}$ with $\left\vert \psi_{f}\right\rangle =\left\vert
N\right\rangle $ for $t\rightarrow\infty$, when the field is zero. The result
of perturbation theory for $C_{N}$ to the lowest order in the control field
$E\left(  t\right)  $ is
\begin{equation}
C_{N}=i^{N}\prod_{k=1}^{N}\mu_{k}\int_{-\infty}^{\infty}dt_{N}E(t_{N}%
)e^{i\bar{\omega}_{N}t_{N}}\!\!\!\int_{-\infty}^{t_{N}}dt_{N-1}E(t_{N-1}%
)e^{i\bar{\omega}_{N-1}t_{N-1}}\ldots\int_{-\infty}^{t_{2}}dt_{1}%
E(t_{1})e^{i\bar{\omega}_{1}t_{1}}\text{,} \label{time_domain}%
\end{equation}
where
\begin{equation}
\bar{\omega}_{n}=\varepsilon_{n}-\varepsilon_{n-1}\text{, }n=1,\cdots
,N\text{,}%
\end{equation}
are the transition frequencies. Utilizing the Fourier transform of the field
\begin{subequations}
\begin{align}
E\left(  t\right)   &  =\frac{1}{2\pi}\int_{-\infty}^{\infty}f\left(
\Omega\right)  e^{-i\Omega t}d\Omega\text{,}\\
f\left(  \Omega\right)   &  =\int_{-\infty}^{\infty}E\left(  t\right)
e^{i\Omega t}dt
\end{align}
in Eq. (\ref{time_domain}) produces
\end{subequations}
\begin{align}
C_{N}  &  =(\frac{i}{2\pi})^{N}\prod_{k=1}^{N}\mu_{k}\int d\Omega_{1}\cdots
d\Omega_{N}f(\Omega_{1})\cdots f\left(  \Omega_{N}\right) \nonumber\\
&  \int_{-\infty}^{\infty}dt_{N}e^{i\left(  \bar{\omega}_{N}-\Omega
_{N}\right)  t_{N}}\!\!\!\int_{-\infty}^{t_{N}}dt_{N-1}e^{i\left(  \bar
{\omega}_{N-1}-\Omega_{N-1}\right)  t_{N-1}}\ldots\int_{-\infty}^{t_{2}}%
dt_{1}e^{i\left(  \bar{\omega}_{1}-\Omega_{1}\right)  t_{1}}\text{.}
\label{CNFour}%
\end{align}
In the above equation, the definite integrals over the frequencies $\Omega
_{k}$ run from $-\infty$ to $\infty$. To calculate the time integral in Eq.
(\ref{CNFour}), we introduce a small imaginary part $\epsilon_{k}%
\rightarrow+0$ in the frequencies,%
\begin{equation}
\Omega_{k}\rightarrow\Omega_{k}+i\varepsilon_{k},k=1,\cdots,N-1\text{.}
\label{img}%
\end{equation}
It is seen from Eq. (\ref{time_domain}) that such a change will not affect the
result, since $E(t)\rightarrow0$ for $t\rightarrow-\infty$; we assume that
$E(t)$ decays at least exponentially for $t\rightarrow\pm\infty$, which is
physically reasonable. Evaluating the integrals in Eq. (\ref{CNFour}) with
respect to $t_{1}$, $t_{2}$, $\cdots$, $t_{N-1}$ yields%
\begin{align}
C_{N}  &  =(\frac{i}{2\pi})^{N}\prod_{k=1}^{N}\mu_{k}\int d\Omega_{1}\cdots
d\Omega_{N}f(\Omega_{1})\cdots f\left(  \Omega_{N}\right) \nonumber\\
&  \int_{-\infty}^{\infty}dt_{N}e^{-i\left(  \Omega_{N}-\bar{\omega}%
_{N}\right)  t_{N}}\frac{\exp\left[  -i\sum\nolimits_{q=1}^{N-1}\left(
\Omega_{q}-\bar{\omega}_{q}\right)  t_{N}\right]  }{\prod\nolimits_{j=1}%
^{N-1}\left[  -i\sum_{p=1}^{j}\left(  \Omega_{p}-\bar{\omega}_{p}\right)
\right]  }\text{,} \label{CN1}%
\end{align}
where all $\Omega_{k}$, $k=1,\cdots,N-1$, carry a hidden small positive
imaginary part as Eq. (\ref{img}). Integrating Eq. (\ref{CN1}) again with
respect to $t_{N}$ and $\Omega_{N}$ produces $C_{N}$ in the frequency
representation,%
\begin{align}
C_{N}  &  =\frac{i\prod_{k=1}^{N}\mu_{k}}{\left(  -2\pi\right)  ^{N-1}}%
\int\prod_{j=1}^{N-1}\frac{f(\Omega_{j})d\Omega_{j}}{\sum\nolimits_{p=1}%
^{j}\left(  \Omega_{p}-\bar{\omega}_{p}\right)  }\int f(\Omega_{N}%
)\delta\left(  \sum\limits_{q=1}^{N}\left(  \Omega_{q}-\bar{\omega}%
_{q}\right)  \right)  d\Omega_{N}\nonumber\\
&  =\frac{i\prod_{k=1}^{N}\mu_{k}}{\left(  -2\pi\right)  ^{N-1}}\int
f(\bar{\omega}_{N}-\sum\nolimits_{q=1}^{N-1}\left(  \Omega_{q}-\bar{\omega
}_{q}\right)  )\prod_{j=1}^{N-1}\frac{f(\Omega_{j})d\Omega_{j}}{\sum_{p=1}%
^{j}\left(  \Omega_{p}-\bar{\omega}_{p}\right)  }\text{.} \label{result}%
\end{align}
An important property arising from Eq.~(\ref{result}) is that for the state
$N$ to be populated by the pulse $E(t)$, the sum of the transition frequencies
should be equal to $\varepsilon_{N}-\varepsilon_{0}$. There is an important
difference with the work of Larsen and Bloembergen\cite{Larsen76254}, in which
the condition $\varepsilon_{N}-\varepsilon_{0}=N\omega$ leads to multiphoton
Rabi oscillations, not to an actual transition, as they have a stationary
periodic field, not a radiation pulse.

We will consider the case where $M,$ the number of the components in the pulse
Eq. (\ref{E0}), is equal to $N,$ the number of the transitions in the
multilevel ladder system, so%
\begin{subequations}
\begin{align}
E\left(  t\right)   &  =s\left(  t\right)  \sum_{l=1}^{N}A_{l}e^{-i\theta_{l}%
}e^{-i\omega_{l}t}+c.c\text{,}\label{ER}\\
f(\omega)  &  =\sum_{l=1}^{N}\left[  A_{l}e^{-i\theta_{l}}S\left(
\omega-\omega_{l}\right)  +A_{l}e^{i\theta_{l}}S\left(  \omega+\omega
_{l}\right)  \right]  \text{,}%
\end{align}
where
\end{subequations}
\begin{equation}
S\left(  \omega\right)  =\int_{-\infty}^{\infty}s\left(  t\right)  e^{i\omega
t}dt\text{,}%
\end{equation}
and each component is only resonant with the corresponding system transition,
\begin{equation}
\left\vert \omega_{k}-\bar{\omega}_{k}\right\vert \ll\left\vert \omega
_{k}-\bar{\omega}_{j\neq k}\right\vert \text{, }k,j=1,\cdots,N
\end{equation}
with the function $S\left(  \omega\right)  $ assumed to be smooth, $\left\vert
dS/d\omega\right\vert \ll\left\vert S\right\vert /\omega_{k}$. The latter
condition means that the typical duration of the pulse significantly exceeds
the transition periods $2\pi/\omega_{k}$, $k=1,\cdots,N$. Then the nearly
resonant terms in Eq. (\ref{result}) are kept,%
\begin{subequations}
\begin{align}
C_{N}  &  \approx\frac{i\prod_{k=1}^{N}\mu_{k}e^{-i\theta_{k}}}{\left(
-2\pi\right)  ^{N-1}}\int A_{N}S(\bar{\omega}_{N}-\omega_{N}-\sum
\limits_{q=1}^{N-1}\left(  \Omega_{q}-\bar{\omega}_{q}\right)  )\prod
_{j=1}^{N-1}\frac{A_{j}S(\Omega_{j}-\omega_{j})d\Omega_{j}}{\sum
\nolimits_{p=1}^{j}\left(  \Omega_{p}-\bar{\omega}_{p}\right)  }\\
&  =\tilde{C}_{N}\prod_{k=1}^{N}\mu_{k}A_{k}e^{-i\theta_{k}}\text{,}
\label{CNAP}%
\end{align}
where%
\end{subequations}
\begin{equation}
\tilde{C}_{N}=\frac{i}{\left(  -2\pi\right)  ^{N-1}}\int S(\bar{\omega}%
_{N}-\omega_{N}-\sum\limits_{q=1}^{N-1}\left(  \Omega_{q}-\bar{\omega}%
_{q}\right)  )\prod_{j=1}^{N-1}\frac{S(\Omega_{j}-\omega_{j})d\Omega_{j}}%
{\sum\nolimits_{p=1}^{j}\left(  \Omega_{p}-\bar{\omega}_{p}\right)  }\text{.}
\label{CNG}%
\end{equation}
If each component of the pulse is exactly resonant with a transition of the
system,%
\begin{equation}
\omega_{k}=\bar{\omega}_{k}\text{,}%
\end{equation}
then a simple form for $\tilde{C}_{N}$,%
\begin{equation}
\tilde{C}_{N}=\frac{i^{N}\tau^{N}}{N!}\text{,} \label{CN0}%
\end{equation}
may be attained with
\begin{equation}
\tau=S\left(  0\right)  =\int_{-\infty}^{\infty}s\left(  t\right)  dt
\label{Effduration}%
\end{equation}
being the effective pulse duration. It's easier to prove Eq. (\ref{CN0}) in
the time domain than in the frequency domain. Considering the control field in
Eq. (\ref{ER}), whose each component is resonant with a particular transition,
the neglect of all nonresonant terms in the Eq. (\ref{time_domain}) yields%
\begin{subequations}
\begin{align}
C_{N}  &  \approx i^{N}\prod_{k=1}^{N}\mu_{k}A_{k}e^{-i\theta_{k}}%
\int_{-\infty}^{\infty}dt_{N}s(t_{N})\int_{-\infty}^{t_{N}}dt_{N-1}%
s(t_{N-1})\ldots\int_{-\infty}^{t_{2}}dt_{1}s(t_{1})\\
&  =i^{N}\prod_{k=1}^{N}\mu_{k}A_{k}e^{-i\theta_{k}}\frac{\left[
\int_{-\infty}^{\infty}s\left(  t\right)  dt\right]  ^{N}}{N!}\\
&  =\frac{i^{N}\tau^{N}}{N!}\prod_{k=1}^{N}\mu_{k}A_{k}e^{-i\theta_{k}%
}\text{,}%
\end{align}
from which Eq. (\ref{CN0}) follows.

If the pulse is not exactly resonant%
\end{subequations}
\begin{equation}
\delta_{k}=\omega_{k}-\bar{\omega}_{k}\neq0\text{,} \label{detuning}%
\end{equation}
then $\tilde{C}_{N}$ is not so simple, except for the case where $\delta
_{k}=\delta$ is independent of $k$, i.e., the detuning is the same for all
frequencies. In this case Eq. (\ref{CN0}) still applies, but Eq.
(\ref{Effduration}) should be modified to become%
\begin{equation}
\tau=S\left(  -\delta\right)  =\int_{-\infty}^{\infty}s\left(  t\right)
e^{-i\delta t}dt\text{.} \label{EDuNot0}%
\end{equation}
Introducing the following change of variable in Eq. (\ref{CNG}),
\begin{equation}
z_{j}=\Omega_{j}-\omega_{j}\text{,} \label{z}%
\end{equation}
yields
\begin{equation}
\tilde{C}_{N}=\frac{i}{\left(  -2\pi\right)  ^{N-1}}\int S(-\sum
\nolimits_{q=1}^{N-1}z_{q}-\Delta_{N})\prod_{j=1}^{N-1}\frac{S(z_{j})dz_{j}%
}{\sum_{p=1}^{j}z_{p}+\Delta_{j}}\text{,} \label{fre_domain}%
\end{equation}
where $\Delta_{k}$ is the cumulant detuning,
\begin{equation}
\Delta_{k}=\sum_{p=1}^{k}\delta_{p},k=1,\cdots,N\text{.} \label{Delta}%
\end{equation}
Inserting delta functions in Eq. (\ref{fre_domain}) produces%
\begin{align}
\tilde{C}_{N}  &  =\frac{i}{\left(  -2\pi\right)  ^{N-1}}\prod_{k=1}%
^{N-1}\left(  \int_{-\infty}^{\infty}dz_{k}\int_{-\infty}^{\infty}%
dz_{k}^{\prime}\right)  S(-\sum\nolimits_{q=1}^{N-1}z_{q}-\Delta_{N}%
)\prod_{j=1}^{N-1}\frac{S(z_{j})\delta\left(  z_{j}-z_{j}^{\prime}\right)
}{z_{j}^{\prime}+\sum_{p=1}^{j-1}z_{p}+\Delta_{j}}\nonumber\\
&  =\frac{\left(  -1\right)  ^{N-1}i}{\left(  2\pi\right)  ^{2\left(
N-1\right)  }}\prod_{k=1}^{N-1}\left(  \int_{-\infty}^{\infty}dz_{k}%
\int_{-\infty}^{\infty}dz_{k}^{\prime}\int_{-\infty}^{\infty}d\tau_{k}\right)
S(-\sum\limits_{q=1}^{N-1}z_{q}-\Delta_{N})\prod_{j=1}^{N-1}S(z_{j}%
)e^{i\tau_{j}z_{j}}\frac{e^{-i\tau_{j}z_{j}^{\prime}}}{z_{j}^{\prime}%
+\sum_{p=1}^{j-1}z_{p}+\Delta_{j}}\text{.}%
\end{align}
From Eq. (\ref{img}) it follows that there is a small positive imaginary part
in each variable $z_{k}^{\prime}$ in Eq. (\ref{z}), so the integration over
$z_{k}^{\prime}$, $k=1,\cdots,N-1$ yields%
\begin{equation}
\tilde{C}_{N}=\frac{i^{N}}{\left(  2\pi\right)  ^{N-1}}\prod_{k=1}%
^{N-1}\left(  \int_{0}^{\infty}d\tau_{k}\int_{-\infty}^{\infty}dz_{k}\right)
G(-\sum\limits_{q=1}^{N-1}z_{q}-\Delta_{N})\prod_{j=1}^{N-1}G(z_{j}%
)e^{i\tau_{j}\left(  \sum_{p=1}^{j}z_{p}+\Delta_{j}\right)  }\text{.}
\label{CNtilde}%
\end{equation}
The same result can also be obtained directly from Eq. (\ref{CNFour}) by
changing from $t_{1},\cdots,t_{N-1}$ to $\tau_{1}=t_{1},\tau_{2}=t_{2}%
-t_{1},\cdots,\tau_{N}=t_{N}-t_{N-1}$ (or $t_{k}=\tau_{1}+\cdots+\tau_{k},$
$k=1,\cdots,N$) and from $\Omega_{j}$ to $z_{j}$ given by Eq. (\ref{z}). As
illustrations of the general formulation above, Gaussian and rectangular
pulses are considered in the following treatment.

\subsection{Gaussian pulse}

If the pulse envelope of the control field is Gaussian,
\begin{subequations}
\label{Gaussian}%
\begin{align}
s\left(  t\right)   &  =\exp\left(  -\frac{\pi t^{2}}{\tau^{2}}\right)
\text{,}\label{Gs}\\
S\left(  \omega\right)   &  =\tau\exp\left(  -\frac{\omega^{2}}{\sigma^{2}%
}\right)  \text{,} \label{Gf}%
\end{align}
with $\sigma$ being the spectral width of the pulse,
\end{subequations}
\begin{equation}
\sigma=\frac{2\sqrt{\pi}}{\tau}\text{,}%
\end{equation}
then substituting Eq. (\ref{Gf}) into Eq. (\ref{CNtilde}) leads to%
\begin{align}
\tilde{C}_{N}  &  =\frac{i^{N}\tau^{N}}{\left(  2\pi\right)  ^{N-1}}%
\prod_{k=1}^{N-1}\left(  \int_{0}^{\infty}d\tau_{k}\int_{-\infty}^{\infty
}dz_{k}\right)  \exp\left[  -\left(  \sum\limits_{q=1}^{N-1}z_{q}+\frac
{\Delta_{N}}{\sigma}\right)  ^{2}-\sum_{j=1}^{N-1}z_{j}^{2}+\sum_{j=1}%
^{N-1}i\tau_{j}\left(  \sum_{p=1}^{j}z_{j}+\frac{\Delta_{j}}{\sigma}\right)
\right] \nonumber\\
&  =\frac{i^{N}\tau^{N}}{\left(  2\pi\right)  ^{N-1}}\prod_{k=1}^{N-1}\left(
\int_{0}^{\infty}d\tau_{k}\int_{-\infty}^{\infty}dz_{k}\right)  \exp\left[
-\sum_{k,j=1}^{N-1}z_{k}A_{kj}z_{j}+\sum_{j=1}^{N-1}b_{j}z_{j}+i\sum
_{j=1}^{N-1}\tau_{j}\frac{\Delta_{j}}{\sigma}-\frac{\Delta_{N}^{2}}{\sigma
^{2}}\right]  \text{,} \label{CNGau}%
\end{align}
where the elements of matrix $A$ and vector $b$ are
\begin{subequations}
\begin{align}
\text{\ }A_{kj}  &  =\text{$\boldsymbol{\text{$\boldsymbol{\delta}$}}$}%
_{kj}+1\text{,}\\
b_{j}  &  =i\sum_{p=j}^{N-1}\tau_{p}-2\frac{\Delta_{N}}{\sigma}\text{.}%
\end{align}
Carrying out the Gaussian integrals in Eq. (\ref{CNGau}) with respect to
$z_{k}$, $k=1,\cdots,N-1$ yields%
\end{subequations}
\begin{equation}
\tilde{C}_{N}=\frac{i^{N}\tau^{N}}{\left(  2\pi\right)  ^{N-1}}\sqrt{\frac
{\pi^{N-1}}{\det A}}\prod_{k=1}^{N-1}\left(  \int_{0}^{\infty}d\tau
_{k}\right)  \exp\left[  \frac{1}{4}\sum_{k,j=1}^{N-1}b_{k}\left(
A^{-1}\right)  _{kj}b_{j}+i\sum_{j=1}^{N-1}\tau_{j}\frac{\Delta_{j}}{\sigma
}-\frac{\Delta_{N}^{2}}{\sigma^{2}}\right]  \text{.}%
\end{equation}
It's easy to verify that%
\begin{subequations}
\begin{align}
\left(  A^{-1}\right)  _{kj}  &  =\text{$\boldsymbol{\text{$\boldsymbol{\delta
}$}}$}_{kj}-\frac{1}{N}\text{, }\\
\det A  &  =N\text{.}%
\end{align}
Performing some algebra produces the perturbative solution for the Gaussian
pulse,%
\end{subequations}
\begin{equation}
\tilde{C}_{N}=\frac{i^{N}\tau^{N}}{2^{N-1}\pi^{\frac{N-1}{2}}N^{\frac{1}{2}}%
}e^{-\frac{\Delta_{N}^{2}}{N\sigma^{2}}}\prod_{p=1}^{N-1}\left[  \int
_{0}^{\infty}\exp\left(  -\frac{i}{N\sigma}\tau_{p}D_{p}\right)  d\tau
_{p}\right]  \exp\left\{  -\frac{1}{4N}P_{N}\left(  \vec{\tau}\right)
\right\}  \text{,} \label{CNFinal}%
\end{equation}
where
\begin{subequations}
\begin{align}
D_{k}  &  =k\Delta_{N}-N\Delta_{k}\text{, }k=1,\cdots,N-1\text{,}\\
P_{N}\left(  \vec{\tau}\right)   &  =\sum_{k=1}^{N-1}k\left(  N-k\right)
\tau_{k}^{2}+\sum_{0<k<j<N}2k\left(  N-j\right)  \tau_{k}\tau_{j}\text{.}%
\end{align}

When $D_{k}\neq0$ and $\sigma\rightarrow0$, the asymptotic behavior of
transition probability\cite{Henrici1977},%
\end{subequations}
\begin{equation}
\left\vert C_{N}\right\vert \varpropto\sigma^{N-1}\frac{\exp\left(
-\frac{\Delta_{N}^{2}}{N\sigma^{2}}\right)  }{\prod_{k=1}^{N-1}\left\vert
D_{k}\right\vert }\text{,} \label{CAsymN}%
\end{equation}
is obtained. The above equation shows that the yield exponentially decays with
the sum of the detuning of the individual transitions, $\Delta_{N}$ from Eq.
(\ref{Delta}). This behavior can be easily understood since, as pointed out
previously, the sum of the field frequencies should be equal to the sum of the
transition frequencies of the system, which is the energy conservation
condition. A simple calculation shows that, for all $\left\vert \delta
_{i}\right\vert \gg\sigma$, the most probable way of meeting the energy
conservation constraint leads to $\left\vert \tilde{C}_{N}\right\vert
\propto\exp\left(  -\Delta_{N}^{2}/N\sigma^{2}\right)  $.

\subsection{Rectangular pulse}

In the case that the pulse envelope of the control field is rectangular,
\begin{subequations}
\begin{align}
s\left(  t\right)   &  =\left\{
\begin{array}
[c]{l}%
1\text{,}\ \text{ }0<t<T\text{,}\\
0\text{,}\ \text{ otherwise. \ \ \ \ \ \ \ }%
\end{array}
\right. \label{recT}\\
S(\omega)  &  =\frac{e^{i\omega T}-1}{i\omega}\text{,} \label{recF}%
\end{align}
the substitution of Eq. (\ref{recF}) into Eq. (\ref{fre_domain}) leads to%
\end{subequations}
\begin{equation}
\tilde{C}_{N}=\frac{i}{\left(  -2\pi\right)  ^{N-1}}\int\mathcal{I}_{N-1}%
\prod\limits_{j=1}^{N-2}\frac{e^{iz_{j}T}-1}{iz_{j}}\frac{dz_{k}}{\sum
_{p=1}^{j}z_{p}+\Delta_{j}}\text{,}%
\end{equation}
where $\mathcal{I}_{N-1}$ is an integral with respect to $z_{N-1}$
\begin{align}
\mathcal{I}_{N-1}  &  =\int\frac{\exp\left[  -i\left(  \sum\nolimits_{q=1}%
^{N-1}z_{q}+\Delta_{N}\right)  T\right]  -1}{-i\left[  \sum\nolimits_{q=1}%
^{N-1}z_{q}+\Delta_{N}\right]  }\frac{e^{iz_{N-1}T}-1}{iz_{N-1}}\frac
{dz_{N-1}}{\sum_{p=1}^{N-1}z_{p}+\Delta_{N-1}}\nonumber\\
&  =\int\frac{\exp\left[  -i\left(  \sum\nolimits_{q=1}^{N-1}z_{q}+\Delta
_{N}\right)  T\right]  }{-i\left[  \sum\nolimits_{q=1}^{N-1}z_{q}+\Delta
_{N}\right]  }\frac{-1}{iz_{N-1}}\frac{dz_{N-1}}{\sum_{p=1}^{N-1}z_{p}%
+\Delta_{N-1}}\text{.}%
\end{align}
This equation can be evaluated by the residue theorem\cite{Henrici1977}. After
checking the three poles in the lower half plane $\operatorname{Im}z_{N-1}%
\leq0$, it can be shown that only the residue of the zero pole is needed
because those of the other two poles do not contribute after integrating with
respect to $z_{N-2}$, so we have%
\begin{align}
\mathcal{I}_{N-1}  &  =-2\pi i\text{Res}\left(  z_{N-1}=0\right) \nonumber\\
&  =\frac{2\pi i}{\sum\nolimits_{q=1}^{N-2}z_{q}+\Delta_{N-1}}\frac
{\exp\left[  -i\left(  \sum\nolimits_{q=1}^{N-2}z_{q}+\Delta_{N}\right)
T\right]  }{\sum\nolimits_{q=1}^{N-2}z_{q}+\Delta_{N}}\text{.}%
\end{align}
Integrating with respect to $z_{N-2}$, $z_{N-3}$,$\cdots,z_{2}$, similarly as
above, reduces $C_{N}$ to an integral with respect to only one variable,%

\begin{align}
\tilde{C}_{N}  &  =\frac{\left(  -1\right)  ^{N}i}{2\pi}\int\frac{e^{-i\left(
z_{1}+\Delta_{N}\right)  T}}{\prod_{q=2}^{N}\left(  z_{1}+\Delta_{q}\right)
}\frac{e^{iz_{1}T}-1}{iz_{1}}\frac{dz_{1}}{z_{1}+\Delta_{1}}\nonumber\\
&  =\frac{\left(  -1\right)  ^{N}i}{2\pi}\int\frac{e^{-i\left(  z_{1}%
+\Delta_{N}\right)  T}}{\prod\nolimits_{q=2}^{N}\left(  z_{1}+\Delta
_{q}\right)  }\frac{-1}{iz_{1}}\frac{dz_{1}}{z_{1}+\Delta_{1}}\text{.}%
\end{align}
This result finally produces a compact form for the perturbative solution with
a rectangular pulse:
\begin{equation}
\tilde{C}_{N}=\frac{\left(  -1\right)  ^{N}i}{2\pi}\int\frac{e^{-i\left(
z+\Delta_{N}\right)  T}}{z\prod_{q=1}^{N}\left(  z+\Delta_{q}\right)
}dz\text{.} \label{CRect}%
\end{equation}
\ If all of the transitions are resonant: $\delta_{p}=0$, $p=1,\cdots,N$, then%
\begin{align}
\tilde{C}_{N}  &  \approx\frac{\left(  -1\right)  ^{N}i}{2\pi}\int
\frac{e^{-izT}}{z^{N+1}}dz\nonumber\\
&  =\frac{i^{N}}{N!}T^{N}\text{,}%
\end{align}
which is consistent with Eq. (\ref{CN0}). If all $\Delta_{q},$ $q=1,\cdots,N$,
are different from each other, then from Eq. (\ref{CRect}),%
\begin{equation}
\tilde{C}_{N}=\left(  -1\right)  ^{N}\sum_{q=0}^{N}e^{-i\left(  \Delta
_{N}-\Delta_{q}\right)  T}\prod_{j=0\left(  j\neq q\right)  }^{N}\left(
\Delta_{j}-\Delta_{q}\right)  ^{-1}%
\end{equation}
with $\Delta_{0}=0$. In the case where all of the detunings are the same, we
have $\Delta_{q}=q\delta$, then%
\begin{equation}
\tilde{C}_{N}=\frac{\left(  -1\right)  ^{N}}{N!}\delta^{-N}\left(
e^{-iT\delta}-1\right)  ^{N}\text{.} \label{CNRectSameD}%
\end{equation}
This expression agrees with Eq. (\ref{CN0}) derived earlier for a pulse of an
arbitrary shape.

An interesting consequence of Eq. (\ref{CNRectSameD}) is that the transition
probability is an oscillating function of $T\delta$,
\begin{equation}
\left\vert \tilde{C}_{N}\right\vert ^{2}=\frac{2^{2N}}{\left(  N!\right)
^{2}}\delta^{-2N}\sin^{2N}\frac{T\delta}{2}\text{.} \label{RecNoise}%
\end{equation}
In particular, for $T\delta=2n\pi$, $n=0,1,\cdots$, $\left\vert \tilde{C}%
_{N}\right\vert ^{2}$ becomes zero. Such an antiresonance is a result of
destructive interference, which eliminates transitions to higher states.

\section{Cooperation between A Weak Field and Noise for a Multi-State Ladder
System}

From Eq. (\ref{CNAP}), it is evident that, in a weak field driven multilevel
ladder system, the transition probability is independent of the phases of the
individual pulse components, but is sensitive to their amplitudes and
frequencies. This section considers the noise-averaged transition probability
over the amplitudes and frequencies of a weak field, and shows that the
population transfer can be enhanced under suitable conditions.

\subsection{Noise in the amplitudes of the control field}

From Eq. (\ref{CNAP}), it is easy to determine the dependence of the control
yield on the amplitudes of the field pulses,%
\begin{equation}
O\left[  E(t)\right]  =\left\vert C_{N}\right\vert ^{2}\approx\alpha_{N}^{2}%
{\displaystyle\prod\limits_{l=1}^{N}}
A_{l}^{2}\text{,}%
\end{equation}
where
\begin{equation}
\alpha_{N}=\left\vert \tilde{C}_{N}\prod_{k=1}^{N}\mu_{k}\right\vert
\end{equation}
is independent of the amplitudes. If each amplitude $A_{l}$ is contaminated
with random uniform noise distributed over $\left[  -\gamma_{l},\gamma
_{l}\right]  $, as in Eq. (\ref{ANoise}), then the noise-averaged outcome of
the control process is%

\begin{subequations}
\begin{equation}
\bar{O}\left[  E(t),\vec{\gamma}\right]  =\alpha_{N}^{2}%
{\displaystyle\prod\limits_{l=1}^{N}}
\left\langle A_{l}^{2}\right\rangle \text{,}%
\end{equation}
where $\left\langle A_{l}^{2}\right\rangle $ is the mean square amplitude of
the $l$-th radiation component,%
\end{subequations}
\begin{equation}
\left\langle A_{l}^{2}\right\rangle =\int(A_{l}^{0}+\tilde{x}_{l})^{2}\rho
_{l}^{\left(  A\right)  }\left(  \tilde{x}_{l}\right)  d\tilde{x}_{l}=\left.
A_{l}^{0}\right.  ^{2}+\left\langle \tilde{x}_{l}^{2}\right\rangle \text{.}%
\end{equation}
It is instructive to compare this expression with the outcome from the control
field not having amplitude fluctuations,
\begin{equation}
O^{\left(  0\right)  }\left[  E\left(  t\right)  \right]  =\alpha_{N}^{2}%
{\displaystyle\prod\limits_{l=1}^{N}}
\left.  A_{l}^{0}\right.  ^{2}\text{.}%
\end{equation}
The ratio
\begin{equation}
\bar{O}/O^{\left(  0\right)  }=\prod_{l=0}^{N}\frac{\left\langle A_{l}%
^{2}\right\rangle }{\left.  A_{l}^{0}\right.  ^{2}}%
\end{equation}
becomes appreciable for a large number of states $N$, even when the ratio
$\left\langle A_{l}^{2}\right\rangle /\left.  A_{l}^{0}\right.  ^{2}$ is close
to $1.0$ for each individual transition. For example, if $\left\langle
A_{l}^{2}\right\rangle /\left.  A_{l}^{0}\right.  ^{2}=1+\varepsilon$ with
$\varepsilon\ll1$, then $\bar{O}/O^{\left(  0\right)  }\cong1+N\varepsilon$.

To find the optimal field amplitude, we set $\left\langle A_{l}^{2}%
\right\rangle =\left.  A_{l}^{0}\right.  ^{2}+\left\langle \tilde{x}_{l}%
^{2}\right\rangle $ and minimize the functional%
\begin{equation}
J_{N}\left(  A_{n}^{\left(  0\right)  }\right)  =\left\vert \alpha_{N}^{2}%
{\displaystyle\prod\limits_{l=1}^{N}}
(\left.  A_{l}^{0}\right.  ^{2}+\left\langle \tilde{x}_{l}^{2}\right\rangle
)-O_{T}\right\vert ^{2}+\alpha\sum_{l=0}^{N}\left.  A_{l}^{0}\right.  ^{2}%
\end{equation}
over $A_{l}^{0}$. Assuming that $\left\{  \left\langle \tilde{x}_{l}%
^{2}\right\rangle \right\}  $ are independent of $\left\{  A_{l}^{0}\right\}
$, the optimal value of $A_{l}^{0}$ satisfies%
\begin{equation}
\left.  A_{l}^{0}\right.  ^{2}+\left\langle \tilde{x}_{l}^{2}\right\rangle
=\text{Constant,}%
\end{equation}
independent of $l$. An important consequence is that, for given $\left\{
\left\langle \tilde{x}_{l}^{2}\right\rangle \right\}  $, the contribution from
noise can beneficially serve to decrease the required amplitude $A_{l}^{0}$ of
the optimal control field leading to a given yield, provided that the yield is
small. Cooperation with noise can be extended to modest control yields beyond
the perturbation approximation, as shown in numerical
simulations\cite{Shuang049270}. Although the presence of strong noise can
considerably reduce the coherent nature of the dynamics, modest target yields
can still be reached, including in a very efficient manner. However, when
attempting to achieve high control yields, a different mechanism is generally
operative involving competition between the deterministic portion of the
control field and the noise\cite{Shuang049270}.

\subsection{Noise in the control frequency spectrum}

Here we consider a weak control field with a Gaussian envelope [Eq.
(\ref{Gaussian})] and frequency noise having Gaussian distribution%
\begin{equation}
f_{k}\left(  \delta_{k}\right)  =\frac{1}{d_{k}\sigma\sqrt{\pi}}\exp\left[
-\left(  \frac{\delta_{k}-\bar{\delta}_{k}}{d_{k}\sigma}\right)  ^{2}\right]
\text{,}%
\end{equation}
where the random variables are the noise contaminated detunings%
\begin{equation}
\delta_{k}=\bar{\delta}_{k}+\tilde{z}_{k}=\omega_{k}^{0}+\tilde{z}_{k}%
-\bar{\omega}_{k}%
\end{equation}
with $\left\{  \tilde{z}_{k}\right\}  $ being uncertainties in the laser
frequencies $\left\{  \omega_{k}\right\}  $ in Eq. (\ref{FNoise}). An
additional source of the detuning noise can be from the transition frequencies
$\left\{  \bar{\omega}_{k}\right\}  $ due to the Doppler or phonon-induced
shift in the control medium. Recalling Eq. (\ref{CNFinal}) yields the
noise-averaged transition probability,%
\begin{align}
\left\langle \left\vert C_{N}\right\vert ^{2}\right\rangle  &  =\prod
_{q=1}^{N}\left(  \int_{-\infty}^{\infty}f_{q}\left(  \delta_{q}\right)
d\delta_{q}\right)  \left\vert C_{N}\right\vert ^{2}\nonumber\\
&  =\frac{\tau^{2N}}{\left(  4\pi\right)  ^{N-1}N}\prod_{p=1}^{N-1}\left(
\int_{0}^{\infty}d\tau_{p}\int_{0}^{\infty}d\tau_{p}^{\prime}\right)
L_{N}\left(  \vec{\tau},\vec{\tau}^{\prime}\right)  \exp\left\{  -\frac{1}%
{4N}\left[  P_{N}\left(  \vec{\tau}\right)  +P_{N}\left(  \vec{\tau}^{\prime
}\right)  \right]  \right\}  \text{.}\label{CNNoise}%
\end{align}
Here, $L_{N}$ is%
\begin{align}
L_{N}\left(  \vec{\tau},\vec{\tau}^{\prime}\right)   &  =\prod_{q=1}%
^{N}\left(  \int_{-\infty}^{\infty}f_{q}\left(  \delta_{q}\right)  d\delta
_{q}\right)  \exp\left[  -\frac{2}{N\sigma^{2}}\Delta_{N}^{2}-\frac{i}%
{N\sigma}\sum_{k=1}^{N-1}\left(  \tau_{k}-\tau_{k}^{\prime}\right)
D_{k}\right]  \nonumber\\
&  =\frac{1}{\sigma^{N}\pi^{N/2}\prod_{k=1}^{N}d_{k}}\prod_{q=1}^{N}\left(
\int_{-\infty}^{\infty}d\delta_{q}\right)  \exp\left[  -\frac{1}{\sigma^{2}%
}\sum_{k,j=1}^{N}\delta_{k}B_{kj}\delta_{j}+\sum_{k=1}^{N}\delta_{k}c_{k}%
-\sum_{k=1}^{N}\frac{\bar{\delta}_{k}^{2}}{d_{k}^{2}\sigma^{2}}\right]
\label{Ln1}%
\end{align}
with the elements of the matrix $B$ and the vector $c$ being
\begin{subequations}
\begin{align}
B_{kj} &  =\frac{1}{d_{k}^{2}}\text{$\boldsymbol{\text{$\boldsymbol{\delta}$}%
}$}_{kj}+\frac{2}{N}\text{,}\\
c_{k} &  =\frac{2\bar{\delta}_{k}}{\sigma^{2}d_{k}^{2}}-\frac{i}{\sigma
}\left(  \bar{V}-V_{k}\right)  \text{,}%
\end{align}
and the parameters $V_{k}$ and $\vec{V}$ specified by%
\end{subequations}
\begin{subequations}
\begin{align}
V_{k} &  =\sum_{j=k}^{N-1}\left(  \tau_{j}-\tau_{j}^{\prime}\right)  \text{,
}k=0,\cdots,N-1\text{; }V_{N}=0\text{,}\\
\bar{V} &  =\frac{1}{N}\sum_{k=1}^{N}V_{k}\text{. }%
\end{align}
Integrating Eq. (\ref{Ln1}) with respect to $\delta_{q}$ yields
\end{subequations}
\begin{equation}
L_{N}\left(  \vec{\tau},\vec{\tau}^{\prime}\right)  =\frac{1}{\prod_{k=1}%
^{N}d_{k}}\sqrt{\frac{1}{\det B}}\exp\left[  \frac{\sigma^{2}}{4}\sum
_{k,j=1}^{N}c_{k}\left(  B^{-1}\right)  _{kj}c_{j}-\sum_{k=1}^{N}\frac
{\bar{\delta}_{k}^{2}}{d_{k}^{2}\sigma^{2}}\right]  \text{,}\label{JNB}%
\end{equation}
with the elements of $B^{-1}$ (inverse matrix of $B$) and the determinant of
$B$ being
\begin{subequations}
\begin{align}
\left(  B^{-1}\right)  _{kj} &  =d_{k}^{2}\left(
\text{$\boldsymbol{\text{$\boldsymbol{\delta}$}}$}_{kj}-\frac{2d_{j}^{2}%
}{N\left(  1+2\bar{d}^{2}\right)  }\right)  \text{,}\\
\det B &  =\frac{1+2\bar{d}^{2}}{\prod_{k=1}^{N}d_{k}^{2}}\text{, }\bar{d}%
^{2}=\frac{1}{N}\sum_{k=1}^{N}d_{k}^{2}\text{.}%
\end{align}

The expression for the scaled transition probability is simplified in two
limiting cases. The first case is that of nearly resonant driving, where
$\left\vert \bar{\delta}_{k}\right\vert \ll\sigma d_{k}$. In this case
$c_{k}=-i\sigma^{-1}\left(  \bar{V}-V_{k}\right)  $ and the term proportional
to $\bar{\delta}_{k}^{2}$ in Eq. (\ref{JNB}), can be neglected. However, it is
easy to see that $L_{N}\left(  \vec{\tau},\vec{\tau}^{\prime}\right)  <1$
because of the noise in the frequency spectrum, and $L_{N}$ rapidly decreases
with the increasing noise intensity parameters $d_{k}$. Therefore, the
transition probabilities $\left\langle \left\vert C_{N}\right\vert
^{2}\right\rangle $ with noise are smaller than without noise when the control
field frequencies are reliably tuned to be resonant with the system transition frequencies.

The role of the noise is reversed in the case of comparatively strong
detuning, $\left\vert \bar{\delta}_{k}\right\vert \gg\sigma d_{k}$. In this
case we have $c_{k}\approx2\bar{\delta}_{k}/\sigma^{2}d_{k}^{2}$, then Eqs.
(\ref{CNNoise}) and (\ref{JNB}) produce%
\end{subequations}
\begin{equation}
\left\langle \left\vert C_{N}\right\vert ^{2}\right\rangle \propto\exp\left[
-\frac{4\bar{\Delta}_{N}^{2}}{N\sigma^{2}\left(  1+2\bar{d}^{2}\right)
}\right]  \label{NoisyAsym}%
\end{equation}
with $\bar{\Delta}_{N}=\sum_{k=1}^{N}\bar{\delta}_{k}$. In this region, it
follows from Eq. (\ref{NoisyAsym}) that increasing the noise intensity
parameter $\bar{d}^{2}$ leads to an exponentially strong increase in the
transition probability. This finding shows how noise can play a constructive
role in a controlled quantum system. This result may be understood from the
fact that for strong detuning, spectral noise can lead to some pulses actually
being closer to resonance. For these pulses the transition probability is
exponentially higher than for nonresonant pulses. As a result, the noise
averaged transition probability is also strongly increased.

Noise-induced enhancement of the transition probability occurs also for
rectangular laser pulses. This is most easy to see when the detunings of all
the frequency components in the pulse are the same and the scaled transition
probability is given by Eq. (\ref{RecNoise}). As noted earlier, the transition
amplitude into the target state is completely eliminated if the detuning
satisfies $\delta=2n\pi/T$. Pulse-to-pulse variation of $\delta$, or
pulse-to-pulse variation of the duration $T$ will suppress the antiresonance
and lead to a nonzero transition probability even when $\left\vert \tilde
{C}_{N}\right\vert ^{2}=0$ for the average values $\delta=\bar{\delta}$,
$T=\bar{T}$. If the width of the distribution over $\delta$ is small compared
to $\bar{\delta}$, but largely exceeds $\pi/T$, then from Eq. (\ref{RecNoise})
we have $\left\langle \left\vert \tilde{C}_{N}\right\vert ^{2}\right\rangle
\approx\frac{\left(  2N\right)  !}{\left(  N!\right)  ^{4}}\bar{\delta}^{-2N}$.

\section{\bigskip Conclusions}

\bigskip This paper explores the dynamics of population transfer in a
multistate ladder quantum system driven by noisy control pulses, with
particular emphasis on identifying circumstances when cooperation between the
field and noise may occur. The noise enters as run-to-run uncertainties in the
control amplitudes, phases and frequencies with the observation being an
ensemble average over many runs as is commonly done in the laboratory. If the
rotating wave approximation is valid, the quantum dynamics in the weak field
limit is greatly simplified and independent of the control phases.
Furthermore, if the objective yield is modest, the control field can cooperate
with amplitude noise to reduce the applied fluence. Frequency noise in the
control field is shown to be capable of enhancing the transition probability
when the detuning is large. In the laboratory implementation of closed-loop
control, the optimal field will be deduced to extract as much beneficial value
as possible from the presence of noise. This paper presents a theoretical
foundation showing that ample opportunity exists to take advantage of noise.

The above conclusions are fully consistent with recent numerical
simulations\cite{Shuang049270}. Although the analytical treatment in this
paper only applies to ladder-configuration systems, the basic conclusions on
the prospects for cooperating with noise should be applicable for optimal
control of many multistate systems. This point is confirmed with a non-ladder
system\cite{Shuang049270} where a high degree of cooperation was found between
the noise and the deterministic part of the field. Regardless of whether it is
dynamically beneficial to fight or cooperate with noise, the optimal field
will appropriately emphasize the dynamical pathways that correspondingly
either work with or circumvent the influence of the noise\cite{Shuang049270}.

\ 

\begin{acknowledgments}
The authors acknowledge support from the Department of Energy and NSF-PHY 0555346.
\end{acknowledgments}

\bibliographystyle{apsrev}

\end{document}